\documentclass[pra,aps,superscriptaddress,nofootinbib,twocolumn]{revtex4-1}
\usepackage{qcircuit}
\usepackage[dvips]{graphicx}
\usepackage{amsmath,amssymb,amsthm,mathrsfs,amsfonts,dsfont}
\usepackage{subfigure, epsfig}
\usepackage{braket}
\usepackage{bm}
\usepackage{enumerate}
\usepackage{color}

\newcommand{\tr}{\mathrm{Tr}}

\begin{document}
\title{Variational circuit compiler for quantum error correction}
\begin{abstract}
Quantum error correction is vital for implementing universal quantum computing. 
A key component is the encoding circuit that maps a product state of physical qubits into the encoded multipartite entangled logical state. 
Known methods are typically not 'optimal' either in terms of the circuit depth (and therefore the error burden) or the specifics of the target platform, i.e. the native gates and topology of a system. 
This work introduces a variational compiler for efficiently finding the encoding circuit of general quantum error correcting codes with given quantum hardware. Focusing on the noisy intermediate scale quantum regime, we show how to systematically compile the circuit following an optimising process seeking to minimise the number of noisy operations that are allowed by the noisy quantum hardware or to obtain the highest fidelity of the encoded state with noisy gates. We demonstrate our method by deriving novel encoders for logic states of the five qubit code and the seven qubit Steane code. We describe ways to augment the discovered circuits with error detection. Our method is applicable quite generally for compiling the encoding circuits of quantum error correcting codes.  
\end{abstract}

\author{Xiaosi Xu}
\email{xiaosi.xu@materials.ox.ac.uk}

\author{Simon C.~Benjamin}
\email{simon.benjamin@materials.ox.ac.uk}

\author{Xiao Yuan}
\email{xiao.yuan.ph@gmail.com}
\affiliation{Department of Materials, University of Oxford, Parks Road, Oxford OX1 3PH, United Kingdom}
\affiliation{Center on Frontiers of Computing Studies, Department of Computer Science, Peking University, Beijing 100871, China}

\maketitle

\section{Introduction}
Quantum error correction is the key for building large-scale universal quantum computers~\cite{FaultTolerance,Shorcode95,CSScode} and is important for mitigating errors in noisy-intermediate-scaled-quantum (NISQ) computing~\cite{mcardle_error-mitigated_2019,bonet2018low,mcclean2019decoding}. By encoding the logical state as a multipartite entangled state of several physical qubits, it allows us to detect and even further correct errors of the physical qubit without destroying the logical state. 
The five qubit `perfect' code is one of the earliest known codes~\cite{5qubitcodetheory96,laflamme1996perfect}. Being the smallest code that is capable of correcting an arbitrary physical error, the code is also advantageous with low-weight stabiliser operators. An even earlier example is the seven qubit Steane code~\cite{steane1996error}. Also known as the smallest 2D colour code, it has low-weight and symmetrical $X$ and $Z$ stabilisers. As a CSS code it further allows transversal Clifford operations, thus the logical gates of the entire Clifford group are inherently fault-tolerant.

In experiment, quantum error detecting and correcting codes  have been implemented in different platforms ranging from superconducting circuits~\cite{12Schoelkopf,15Martinis,15Chow,15diCarlo,gong2019experimental}, trapped ions~\cite{chiaverini2004realization,11Blatt, 14Blatt,linke2017fault}, NV centers in a diamond~\cite{taminiau2014universal}, optics~\cite{pittman2005demonstration,lu2008experimental,yao2012experimental,bell2014experimental}, and others~\cite{waldherr2014quantum,cramer2016repeated}. However, those experiments are limited to handling only a certain type of errors or with a particular logical state. Comprehensive demonstration of full error correction remains a challenge for the field, mainly due to experimental imperfect controls of the physical qubits and theoretical inefficient compiling of the encoding and decoding processes. Achieving proof-of-principle realisations using near-term noisy quantum devices is therefore difficult, as the encoding, parity checks and decoding processes may include several dozens of imperfect gates as well as non-trivial environmental decoherence, so that the error burden may go beyond the capability that a code can correct and therefore lead to a logical error that cannot be detected and corrected. 

Therefore, realising error correction with NISQ hardware requires both theoretical and experimental advances. Theoretically, the use of classical approaches to reinforce the performance of error correction codes has been exploited~\cite{johnson2017qvector,nautrup2018optimizing}. From the experimental perspective, the manipulation of qubits should be improved to the highest accuracy. However, different physical systems have different hardware arrangements, control pulses, native types of multi-qubit gates, etc. For example, the natural entangling operation between two qubits with certain superconducting circuits and NV centers is a form of CNOT or CPhase gate~\cite{barends2014superconducting,dicarlo2010preparation,wang2013universal,pfaff2014unconditional}, while it is a M\o lmer-S\o rensen gate for many trapped ion systems~\cite{monz2009realization,tan2015multi}, and sqrt-SWAP gate may be the native operation in quantum dot systems~\cite{cai2019silicon,buonacorsi2018network}.
Meanwhile, conventional methods for the realisation of error correcting codes do not necessarily involve rigorous analytic or numerical optimisation and therefore may have an unnecessarily large number of gates. It is thus theoretically important to more efficiently compile quantum error correcting protocols against a specific hardware target. 

There are a number of approaches to quantum compilation documented in the literature. These range from classical methods based on exploiting perfect circuit isomorphisms (e.g. gate commutation)~\cite{prasad2006data,maslov2008quantum,amy2014polynomial,heyfron2018efficient,nam2018automated}, to methods that can `discover' near-equivalent circuits using a quantum device for the compilation (or, for small circuits, an emulator~\cite{jones2018quantum,heya2018variational,khatri2019quantum,sharma2019noise}. In this work, we adopt an approach more closely related to the latter, since we compile in such a fashion as to support any given noisy quantum hardware where formal circuit equivalence may be difficult or impossible to determine. 
We construct a variational compiler to automatically search for circuits that encode a target logical state of an error correcting code. The procedures we describe aim to seek circuits that are optimised in a defined way, such that they meet all the constraints under the particular condition and they are most favourable by design. For example, the `optimised' circuit can be one that has the minimum number of two-qubit gates. Our procedures cannot guarantee that the best solution found is the strictly optimal solution. However we expect that this may indeed be the case when our solution is more efficient than the best circuits previously reported in the literature, or when the automated process `re-discovers' a previously known compact solution.

Our compiler first maps the problem into a ground state searching problem where the desired logical state is the ground state of the given (synthetic) Hamiltonian. Different from the conventional variational quantum eigensolver cases where the energy spectra are unknown, here the energy spectra are known and can be altered.
Next, we consider a set of parameterised ansatz circuits and make use of the variational imaginary time evolution method~\cite{mcardle2018variational,yuan2018theory} to find the ground state, thus discovering the encoding circuit. 
The ansatz circuit can be tailored to meet specific requirements of any hardware system and any optimisation target. For example, when considering quantum hardware with only single- and two-qubit gates, we might either minimise the number of two-qubit gates or minimise the overall infidelity of the prepared state. 

The structure of the paper is as follows. In Sec.~\ref{Sec:QEC}, we first review the framework of quantum error correction and introduce the problem of circuit compiling. In Sec.~\ref{Sec:compiling}, we introduce the variational compiling algorithm including the construction of the Hamiltonian, the design of ansatz, and the variational imaginary time evolution. In Sec.~\ref{sec:numerical}, we show numerical realisations of the compiling algorithm for different logical states of the five qubit and seven qubit codes. We consider both the noise-free and noisy circuits, and for the latter case, we show the fidelity of the prepared state can be boosted through different levels of post selection. We compare our results to existing circuits and discuss its applicability in NISQ computing. We discuss applications of our results and summarise in Sec.~\ref{sec:discussion}.

\section{Background: quantum error correction}\label{Sec:QEC}
The main idea of quantum error correction is to encode  logical qubit(s) with a greater number of noisy physical qubits. Many quantum error correcting codes can be described by the stabiliser formalism, where the code space is determined by the joint positive eigenspace of a set of commutative stabiliser operators. Specifically, considering the Pauli group $G_{n}$ on $n$ qubits, 
\begin{equation}
G_{n} \equiv\{ \pm I, \pm i I, \pm X, \pm i X, \pm Y, \pm i Y, \pm Z, \pm i Z\}^{\otimes n},
\end{equation}
the set of stabilisers $S$ for a quantum error correcting code is a subset of $G_n$ such that $-I\notin S$ and elements in $S$ commute with each other. Suppose $S$ is generated by $G=\left\langle g_{1}, \ldots, g_{l}\right\rangle$, then the code space corresponds to quantum states $\ket{\Phi}_L$ satisfying $g_i\ket{\Phi}_L=\ket{\Phi}_L$ for all stabilisers $g_i$.
Here we introduce two of the most well known small quantum  error correcting codes---the five qubit perfect code and the seven qubit Steane code, which encode one logical qubit state with five and seven physical qubits, respectively. 
A stabiliser set for the five qubit code is 
$$\{XZZXI, IXZZX, XIXZZ, ZXIXZ\},$$ and for the Steane code is 
\begin{equation}\nonumber
    \begin{aligned}
\{IIIXXXX,XXIIXXI,XIXIXIX, \\
IIIZZZZ,ZZIIZZI,ZIZIZIZ\}.
\end{aligned}
\end{equation}

In the logical subspace, the code is further uniquely determined by the logical $Z_L$ operator, with $Z_L\ket{0}_L = \ket{0}_L$ and $Z_L\ket{1}_L = -\ket{1}_L$. A general logical pure state can be thus represented as $\ket{\Phi}_L=a\ket{0}_L+b\ket{1}_L$. For the five and seven qubit codes, all the logical Pauli operators can be transversely realised with corresponding identical local Pauli operators.  That is, we have $Z_L=ZZZZZ$ and $Z_L=ZZZZZZZ$ for the five and seven qubit codes, respectively. For example, the logical $\ket{0}_L$ and $\ket{1}_L$ states of the five qubit code are defined by
\begin{equation} \label{QSS:Fid:logic-qbit1}
\begin{aligned}
\ket{0}_L=&\frac{1}{4}[\ket{00000}-\ket{00011}+\ket{00101}-\ket{00110}\\
&+\ket{01001}+\ket{01010}-\ket{01100}-\ket{01111}\\
&-\ket{10001}+\ket{10010}+\ket{10100}-\ket{10111}\\
&-\ket{11000}-\ket{11011}-\ket{11101}-\ket{11110}],\\
\ket{1}_L=&\frac{1}{4}[-\ket{00001}-\ket{00010}-\ket{00100}-\ket{00111}\\
&-\ket{01000}+\ket{01011}+\ket{01101}-\ket{01110}\\
&-\ket{10000}-\ket{10011}+\ket{10101}+\ket{10110}\\
&-\ket{11001}+\ket{11010}-\ket{11100}+\ket{11111}].\\
\end{aligned}
\end{equation}

In this work, we focus on the problem of how to prepare a logical quantum state $\ket{\Phi}_L=a\ket{0}_L+b\ket{1}_L$ by applying an encoding circuit to a given easily-prepared initial state.  

We aim to find suitable circuits automatically and in a fashion that is more optimal according to some user-specified criteria. 
As context for our work, we now exhibit a few previously-reported examples of encoding circuits for the five qubit code and the Steane code.

\begin{figure}[t]
\centering
\includegraphics[width=0.55\columnwidth]{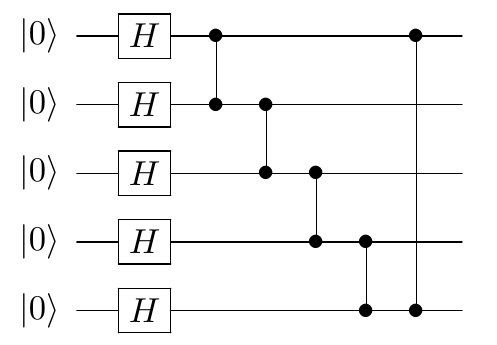}\\(a)\\
\includegraphics[width=0.9\columnwidth]{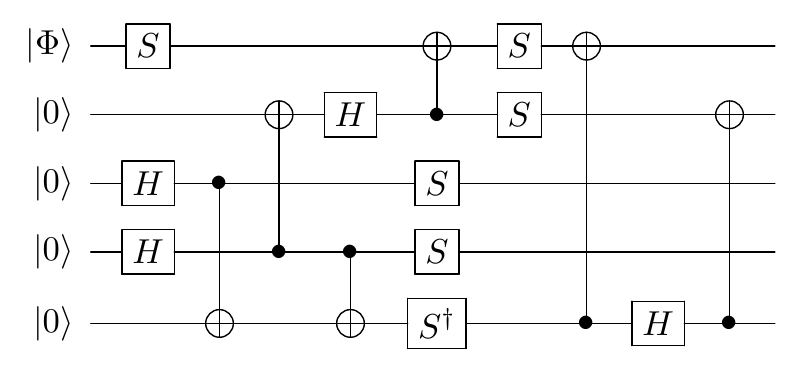}\\
(b)
\caption{Encoding circuit for the five qubit code. (a) Circuit to encode the logical minus state $\ket{-}_L$~\cite{xu2018integrity}. (b) Circuit to encode an arbitrary logical state~\cite{gong2019experimental}.}
\label{fig:five-minus}
\end{figure}

For the five qubit code, at minimum five two-qubit gates are required to prepare a logical state. The circuit shown in Fig.~\ref{fig:five-minus}(a) encodes a logical minus state $\ket{-}_L$ with five single-qubit gates and five CPhase gates~\cite{xu2018integrity}. However, to encode an arbitrary logical state, an extra two-qubit gate is required; a suitable circuit is shown in Fig.~\ref{fig:five-minus}(b) and was recently reported in Ref.~\cite{gong2019experimental}.

\begin{figure}[t]
\centering
\includegraphics[width=0.65\columnwidth]{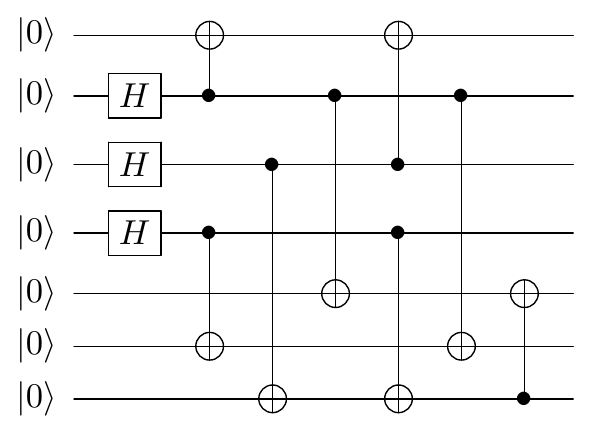}\\
(a)\\
\includegraphics[width=0.9\columnwidth]{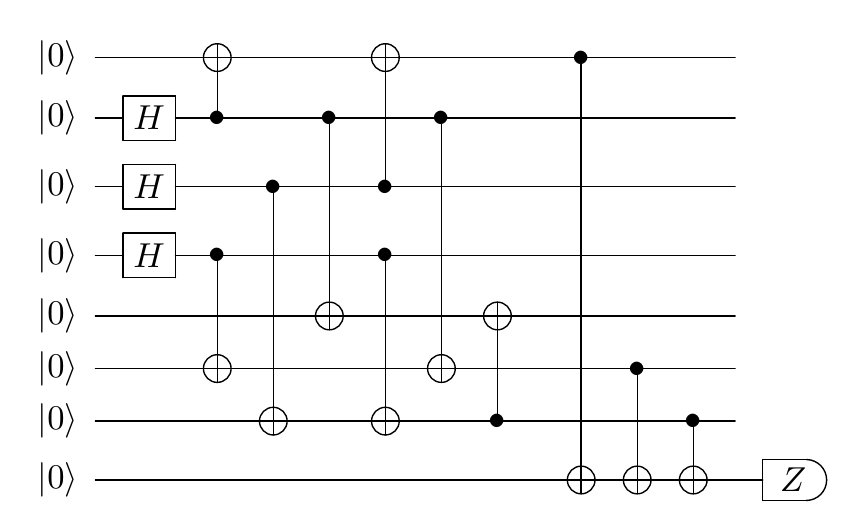}\\
(b)\\
\includegraphics[width=0.85\columnwidth]{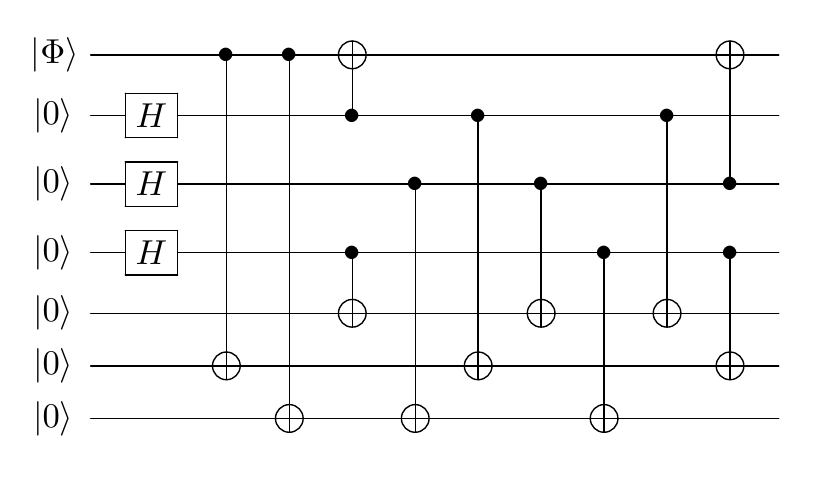}\\
(c)\\
\caption{Previously-known encoding circuits for the seven qubit code. (a) Circuit to encode a logical zero state $\ket{0}_L$\cite{paetznick2011fault}. 
(b) Circuit to fault-tolerantly encode a logical zero state $\ket{0}_L$~\cite{goto2016minimizing}. {Note that logical Clifford eigenstates can also be prepared with this circuit by applying transversal logical Clifford gates after the error detection process}. Three CNOT gates are applied to the ancilla qubit, which is then measured in $\{0,1\}$ basis. If the measurement result is 1, indicating more than one bit-flip errors occurring, the whole circuit is abandoned and restarted from the beginning, until the ancilla qubit is measured to be 0.
(c) Circuit to encode an arbitrary logical state~\cite{xu2018integrity}.}
\label{fig:seven-zero}
\end{figure}

Encoding a logical qubit with the seven qubit Steane code requires more multi-qubit gates. Fig.~\ref{fig:seven-zero}(a) shows the circuit to prepare a logical zero state $\ket{0}_L$ with 8 CNOT gates~\cite{paetznick2011fault}. This circuit can also be fault-tolerant given three additional two-qubit gates and one ancilla qubit~\cite{goto2016minimizing}, as shown in Fig.~\ref{fig:seven-zero}(b). To encode an arbitrary logical state, one may use the circuit shown in Fig.~\ref{fig:seven-zero}(c), which has 11 CNOT gates~\cite{xu2018integrity}.

It is in general non-trivial to find an efficient encoding circuit for a given error correcting code, and the circuit found by hand may not be optimal or compatible with specific experimental hardware. 
The present work solves this problem by introducing a variational way of compiling the encoding circuit. 
We show in the following that even the presented encoding circuits for the five and seven qubit codes are not necessarily optimal in terms of the number of two-qubit gates.

\section{Variational circuit compiling for quantum error correction}\label{Sec:compiling}
Here we introduce the variational circuit compiler for preparing a logical state of an error correcting code. The key idea is to construct a Hamiltonian so that the target logical state is its ground state. Then with a parameterised ansatz circuit, we optimise the parameters in order to find the ground state of the Hamiltonian and hence the encoding circuit of the target logical state. We can either realise the variational circuit compiler with a classical emulator for small error correcting codes, or with a quantum computer for the general case. 
In the following, we first introduce the variational circuit compiler and show how to design the Hamiltonian.  Then we review the recently proposed variational imaginary time evolution for finding the ground state of the Hamiltonian. We also present the realisation of the compiler with quantum circuits. Finally we discuss ansatz design with respect to different quantum hardware. 

\subsection{Variational circuit compiler}
We first show that any logical state of a stabiliser code can be straightforwardly described as the ground state of a Hamiltonian. 
Suppose an error correcting code is stabilised by the generator set $\{g_i\}$. By definition we have 
\begin{equation}
\begin{aligned}
g_i\ket{\Phi}_L=\ket{\Phi}_L,\ \ \ \forall i
\end{aligned}
\end{equation}
where $\ket{\Phi}_L$ is an arbitrary logical state. The stabilisers only force the state into the code space. Suppose the logical state is a single-qubit state, it can be further uniquely determined by an additional logical operator 
\begin{equation}\label{Eq:Odef}
O_L=
\ket{\Phi}_L\bra{\Phi}_L-\ket{\Phi}^{\bot}_L\bra{\Phi}^{\bot}_L.
\end{equation}
Suppose $\ket{\Phi}_L=\alpha\ket{0}_L+\beta\ket{1}_L$, it can be further decomposed as a linear sum of logical Pauli operators $X_L$, $Y_L$, and $Z_L$ as
\begin{eqnarray}
&&O_L=\\
&&(\alpha\beta^*+\alpha^*\beta)X_L-i(\alpha^*\beta-\alpha\beta^*)Y_L-(\beta\beta^*-\alpha\alpha^*)Z_L,\nonumber
\end{eqnarray}
which satisfies $O_L\ket{\Phi}_L=\ket{\Phi}_L$.
In general, when the error correcting code encodes more than one qubit, one can always construct a set of logical operators that determines any logical state. Therefore, we can construct a Hamiltonian, 
\begin{equation}\label{Eq:Hdef}
\begin{aligned}
H=-\sum_ic_ig_i-c_oO_L,
\end{aligned}
\end{equation}
and the target state $\ket{\Phi}_L$ is its unique ground state
\begin{equation}
\begin{aligned}
H\ket{\Phi}_L=-\left(\sum_i c_i+c_o\right)\ket{\Phi}_L,
\end{aligned}
\end{equation}
with energy $E_0=-\left(\sum_i c_i+c_o\right)$. As the terms of the Hamiltonian commute with each other, its eigenstate should be the eigenstate of each term.
Thus, it is not hard to see that the first excited state has an energy $E_1=E_0+2\min\{c_i, c_o\}$.

To find the encoding circuit that prepares the target logical state $\ket{\Phi}_L$, we employ
variational methods for determining a Hamiltonian's ground state (an approach of much current interest). 
We first prepare a trial state via a parameterised quantum circuit, called an ansatz, $\psi(\vec\theta)=V(\vec\theta)\ket{\bar{0}}$, where $\ket{\bar{0}}$ refers to a quantum register of which all the data qubits are initialised at $\ket{0}$, and $V(\vec\theta)$ is described with $m$ parameters, $V(\vec\theta)=V_m(\theta_m)...V_2(\theta_2)V_1(\theta_1)$. Suppose the ground state of the Hamiltonian $H$ can be represented by the circuit ansatz, then the problem is rephrased as finding an set of parameters $\vec\theta_{min}$ which minimises the energy
\begin{equation}
\begin{aligned}
E_{\min}=\min\left\{\langle\psi(\vec\theta)|H|\psi(\vec\theta)\rangle\right\}.
\end{aligned}
\end{equation}
The minimisation can be accomplished by any optimisation algorithm, such as simple gradient descent or the imaginary time evolution~\cite{mcardle2018variational,yuan2018theory} as we presently review. The design of ansatz with respect to different quantum hardware will also be discussed shortly. 

In practice, we may not be able to find the exact ground state, for example because it lies outside the set of states reachable from the ansatz, or because of the existence of gate noise, etc. Suppose the minimal energy we can find is $E_{\min}$, then we can also lower bound the fidelity between the state $\rho$ we find and the target logical state $\ket{\Phi}_L$ according to
\begin{equation}\label{Eq:fidelity}
\begin{aligned}
F = {\braket{\Phi|\rho|\Phi}_L} \ge 1-(E_{\min}-E_0)/c,
\end{aligned}
\end{equation}
where we denote $c=2\min\{c_i, c_o\}$ and assume $E_{\min}\in [E_0, E_1]$. The proof of Eq.~\eqref{Eq:fidelity} can be found in Appendix.  Therefore, when observing an energy that is close to the ground state energy, we are assured that the state is indeed close to the exact ground state. In the rest of this paper, we thus only focus on minimising the energy of the Hamiltonian.

\subsection{Variational simulation with imaginary time evolution}
In previous studies we have found that the imaginary time evolution method can outperform conventional optimisation methods~\cite{mcardle2018variational,yuan2018theory}, therefore we opt to use the variational imaginary time approach as our ground state finding strategy. Needless to say, other variational methods could equivalently be substituted and this is an area for future study. We briefly review the theory here for self-consistency, including both the pure and mixed state cases thus supporting both noiseless and noisy operations in realising the encoding circuit.

\subsubsection{Pure state}
The imaginary time evolution is defined as
\begin{equation}
\begin{aligned}
\ket{\phi(\tau)} = \frac{e^{-H\tau}\ket{\phi(0)}}{\sqrt{\bra{\psi(0)}e^{-2H\tau}\ket{\phi(0)}}}
\end{aligned}
\end{equation}
or equivalently
\begin{equation}
\begin{aligned}
\frac{\partial \ket{\phi(\tau)}}{\partial \tau}=-(H-E_{\tau})\ket{\phi(\tau)},
\end{aligned}
\end{equation}
with $\tau$ being imaginary time and  $E_\tau=\langle\phi(\tau)|H|\phi(\tau)\rangle$. As the amplitudes of all excited eigenstates decay faster than the ground state, we always have $\ket{\phi(\infty)}$ being the ground state of the Hamiltonian $H$. While the imaginary time evolution cannot be directly realised via a unitary quantum circuit, it can be emulated via the variational algorithm.  
Assuming the state $\ket{\phi(\tau)}$ can be well approximated by a parameterised state $\ket{\phi(\tau)}=\ket{\psi(\theta_1,\theta_2,...)}$ with real parameters $\theta_i$, the imaginary time evolution of the quantum state $\ket{\psi(\tau)}$ can be mapped to the evolution of the parameters as
\begin{equation}
\begin{aligned}
\sum_j A_{i,j}\dot{\theta}_i=-B_i,
\end{aligned}
\end{equation}
where 
\begin{equation}\label{Eq:ABdef}
\begin{aligned}
A_{i,j}=&\Re\bigg(\frac{\partial\bra{\psi(\vec{\theta}(\tau))}}{\partial\theta_i}\frac{\partial\ket{\psi(\vec{\theta}(\tau))}}{\partial\theta_j}+\\
&\frac{\partial\bra{\psi(\vec{\theta}(\tau))}}{\partial\theta_i}\ket{\psi(\vec{\theta}(\tau))}\frac{\partial\bra{\psi(\vec{\theta}(\tau))}}{\partial\theta_j}\ket{\psi(\vec{\theta}(\tau))}\bigg),\\
B_i=&\Re\left(\frac{\partial\bra{\psi(\vec{\theta}(\tau))}}{\partial\theta_i}H\ket{\psi(\vec{\theta}(\tau))}\right),
\end{aligned}
\end{equation}
Suppose for the initial state we have $\ket{\phi(\tau)}=\ket{\psi(\theta_1(0),\theta_2(0),...)}$, thus we can update the parameters $\vec\theta=(\theta_1,\theta_2,...)$  via $\vec\theta(\tau+\Delta\tau)=\vec\theta(\tau)+\Delta\tau*\vec{\dot\theta}(\tau)$ to emulate imaginary time evolution. Here we chose $\Delta\tau$ to be a sufficiently small step size. 

\subsubsection{Mixed state}
When the state is a mixed state, the imaginary time evolution obeys~\cite{yuan2018theory}
\begin{equation}
\begin{aligned}
\frac{\partial}{\partial \tau} \rho(\tau)=-(\{H, \rho(\tau)\}-2\langle H\rangle \rho(\tau)). 
\end{aligned}
\end{equation}
When considering a parameterised density matrix $\rho(\vec{\theta}(\tau))$, the imaginary time evolution of $\rho(\vec{\theta}(\tau))$ is mapped to the evolution of the parameters as
\begin{equation}
\begin{aligned}
\sum_{i} C_{j, i} \dot{\theta}_{i}=-D_{j},
\end{aligned}
\end{equation}
where 
\begin{equation}
\begin{aligned}
C_{j, i}&=\operatorname{Tr}\left[\frac{\partial \rho(\vec{\theta}(\tau))}{\partial \theta_{j}} \frac{\partial \rho(\vec{\theta}(\tau))}{\partial \theta_{i}}\right],\\
D_{j}&=\operatorname{Tr}\left[\frac{\partial \rho(\vec{\theta}(\tau))}{\partial \theta_{j}}\{H, \rho(\vec{\theta}(\tau))\}\right].
\end{aligned}
\end{equation}

\subsection{Implementation on quantum circuits}

Our variational circuit compiler can be emulated with a classical computer for small error correcting codes or it can be implemented with quantum circuits for general codes. Here we briefly discuss the implementation of the pure state case with quantum circuits and we refer to Ref.~\cite{yuan2018theory} for the discussion of the mixed state case. 

As the target logical state is the unique ground state of a Hamiltonian, we thus make use of the variational imaginary time evolution method to find the ground state. 
For the pure state case, we need to measure every term of $A$ and $B$ defined in Eq.~\eqref{Eq:ABdef}. Suppose we consider an ansatz $\psi(\vec\theta)=V(\vec\theta)\ket{\bar{0}}$, where each $V_i(\theta_i)$ is a single-qubit rotation around the $X$, $Y$ or $Z$ axis of the Bloch sphere. We can then decompose the derivative of the state as 
\begin{equation}
\begin{aligned}
\frac{\partial\ket{\psi(\vec\theta)}}{\partial\theta_i}=f_i\ket{\varphi_i(\vec\theta)},
\end{aligned}
\end{equation}
where 
$$\ket{\varphi_i(\vec\theta)}=U_i(\vec\theta)\ket{0}=V_m(\theta_m)...\sigma_iV_i(\theta_i)...V_2(\theta_2)V_1(\theta_1)\ket{0}.$$ 
Thus  each term of $A$ consists of 
\begin{equation}
\begin{aligned}
&\frac{\partial\bra{\psi(\vec\theta)}}{\partial\theta_j}\ket{\psi(\vec\theta)}=f_j^*\langle\varphi_j(\vec\theta)|\psi(\vec\theta)\rangle=f_j^*\bra{0}U_j^*(\vec\theta)V(\vec\theta)\ket{0}, \\
&\frac{\partial\bra{\psi(\vec\theta)}}{\partial\theta_j}\frac{\partial\ket{\psi(\vec\theta)}}{\partial\theta_i}=f_if_j^*\bra{0}U_j^*(\vec\theta)U_i(\vec\theta)\ket{0}.
\end{aligned}
\end{equation}
Similarly, each $B_j$ is the real part of
\begin{equation}
\begin{aligned}
\frac{\partial\bra{\psi(\vec\theta)}}{\partial\theta_j}H\ket{\psi(\vec\theta)}=\sum_kf_j^*l_k\bra{0}U_j^*(\vec\theta)H_kV(\vec\theta)\ket{0},
\end{aligned}
\end{equation}
where we assume $H=\sum_k \lambda_kH_k$ with $H_k$ representing a tensor product of Pauli matrices. As all those terms are in a general form of 
$$
a \Re \left( e^{i\theta} \bra{\bar{0}} U \ket{\bar{0}} \right),
$$
they can be efficiently measured with the Hadamard test quantum circuit or equivalent but simpler methods~\cite{li_efficient_2017,mitarai_methodology_2019}. 

\begin{figure}[b]
\begin{centering}
\includegraphics[width=1\columnwidth]{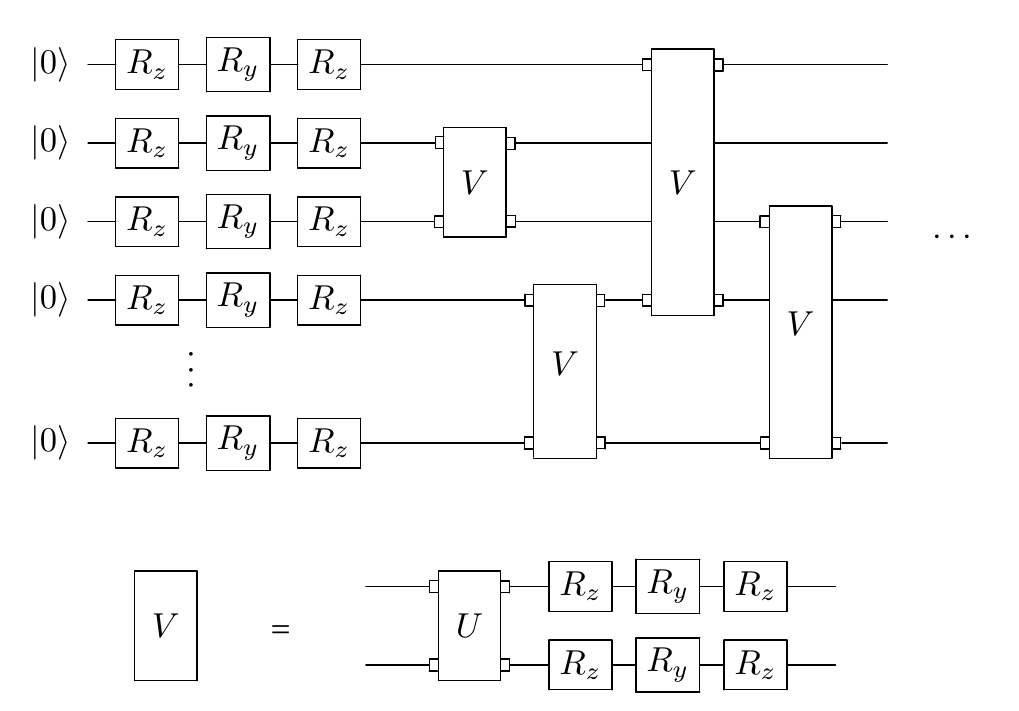}\\
\caption{Building block of the ansatz. The circuit starts with three single-qubit gates applied to each of the data qubits. Then elementary blocks are randomly inserted into the circuit. Each elementary block consists of one two-qubit gate followed with three single-qubit gates on both of the two data qubits. $R_y$ and $R_z$ represent a single-qubit rotation over the $Y$ and $Z$ axes of the Bloch sphere respectively, where the rotation angle is the parameter to be updated over time. }
\label{fig:buildingBlock}
\end{centering}
\end{figure}

\subsection{The ansatz}
Our variational circuit compiler assumes the logical state can be prepared by a parameterised ansatz. For different quantum hardware, the ansatz can have different structures. Here, we introduce the general structure of the  ansatz considered in this work, as shown in Fig.~\ref{fig:buildingBlock}. We consider parameterised single-qubit gates rotating along the Pauli basis. For example, the gate $R_x$ is defined as $R_x = \exp(-i{\theta}X/2)$ with the rotation angle being a variable parameter $\theta$. The definitions of $R_y$ and $R_z$ are similar. We also consider general two-qubit gates composed by a fixed two-qubit gate followed by six parameterised single-qubit gates. 
The overall structure of the ansatz is shown in Fig.~\ref{fig:buildingBlock}, where three single-qubit rotations with different parameters are firstly applied to each of the data qubits following the order of $R_zR_yR_z$. The gate set is chosen such that an arbitrary single-qubit rotation can be realised. With given constraints in the connectivity of the qubits and the type of two-qubit gates that can be realised in a given quantum hardware, a certain number of multi-qubit sets are then inserted into the circuit, where one gate set consists of a multi-qubit gate followed with three single-qubit gates applied to each of the data qubits.

For a given ansatz, which is either randomly generated or following a certain procedure, we update the parameters of the single-qubit gates in order to minimise the energy of the Hamiltonian. Based on different ans\"atze or different initial values of parameters, we could either reach the ground state verified by the ground state energy, or we may reach a local minimum. In the latter case, it may indicate that the ansatz cannot represent the target state, so the experiment is abandoned and restarted until the energy reaches close to the ground state energy.
In order to minimise the number of parameters or single-qubit gates, circuit compilation is applied with the following rule: after each several steps, gates with small parameters are removed; this process continues until the energy starts to increase. The technique cannot guarantee to find a circuit with minimum number of parameters, though many equivalent circuits can be found and selected.

In practice, one may also need to change the structure of the ansatz when the previously selected ansatz is not powerful enough to represent the target state. Different strategies can be applied here by either adding more single and two-qubit gates or fully adopting a different ansatz structure. Trying all possible ansatz structures is in general impossible for a large error correcting code. Therefore one may either systematically explore a given family of ansatz structures, or alternatively we may apply some kind of circuit morphing algorithm to the ansatz (as recently discussed in Ref.~\cite{jones2018quantum,xu2019variational,grimsley2019adaptive}). In the following, we focus on the five- and seven-qubit codes, and show numerical emulation of the variational compiler for these two codes with different ansatz structures.

\section{Numerical simulations}
\label{sec:numerical}

In this section we present numerical simulations for the variational circuit compiler.  The simulation is performed on a classical computer with the Quantum Exact Simulation Toolkit (QuEST) package~\cite{jones2019quest}, which is a high performance classical simulator written in C/C++. We focus on the five and seven qubit codes and consider two scenarios with noiseless and noisy gates. Several particular states are considered in the simulation, including eigenstates of the Pauli basis $\ket{0}_L$, $\ket{-}_L$ and the magic state $\ket{T}_L=(\ket{0}+e^{{\pi}i/{4}}\ket{1})/\sqrt{2}$. For the Hamiltonian, we set the coefficients of all the terms to be the same and normalise them so that the ground state energy is $-1$. In particular, the Hamiltonian corresponding to a logical state $\ket{\Phi}_L$ of the five or seven qubit code is 
\begin{equation}
\begin{aligned}
H = -\frac{1}{n}\left(\sum_ig_i+{O}_L\right),
\end{aligned}
\end{equation}
where $g_i$ are the stabilisers for the five or seven qubit code, ${O}_L$ is the logical operator defined in Eq.~\eqref{Eq:Odef}, $n=5$ for the five qubit code, and $n=7$ for the seven qubit code. We can verify that $H\ket{\Phi}_L=E_0\ket{\Phi}_L$ with $E_0 = -1$ and $E_1 = -(n-2)/n$ for the first excited state. Following Eq.~\eqref{Eq:fidelity}, when we find a state $\rho$ with energy $E_{\min}$, its fidelity to the target state $\ket{\Phi}_L$ is lower bounded by
\begin{equation}
\begin{aligned}
F\ge 1-\frac{n}{2}(E_{\min}+1),
\end{aligned}
\end{equation}
when $E_{\min}\in [-1,-(n-2)/n]$.

We also consider different constraints of the circuit topology for different hardware structures, as revealed in the choice of ans\"atze. The constraints could be from practical experimental limitations or inferred from preferences in a future experimental design. In this paper, we take three constraints as examples to illustrate our method:
\begin{enumerate}[(1)]
    \item Minimise the number of two-qubit gates, as most quantum hardware has a lower fidelity for two-qubit gates;
    \item Only use a single type of two-qubit gate, reflecting the fact that hardware typically supports one entangling process at the physical level;
    \item Only apply nearest-neighbour interactions, such as in superconducting qubit systems.
\end{enumerate}
Note that while searching for circuits satisfying (2) or (3), (1) is also applied by default, as more simplified circuits are usually preferred. At the start of one experiment, an ansatz is generated based on the constraints, with all parameters initialised from a small value around zero. The parameters are then updated through the variational imaginary time algorithm, until the energy becomes static in a local minimum, in which case the ansatz is abandoned, or goes to the ground state energy, in which case we consider the current ansatz is successfully configured. The number of parameters is also gradually reduced in the simulation process: if a certain parameter is found to be around zero, a further simulation is attempted with that gate omitted. This procedure continues until the energy starts to increase. With this trick, we manage to substantially reduce the number of parameters and accelerate the searching process.

In the following, we give several examples of applying our compiler for certain logical states prepared with the five and seven qubit code.

\begin{figure*}[t]
\begin{centering}
\includegraphics[width=0.6\columnwidth]{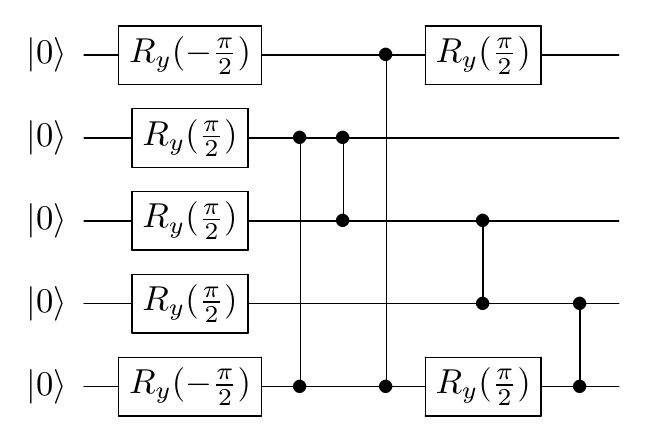}\\
(a)\\
\includegraphics[width=1.6\columnwidth]{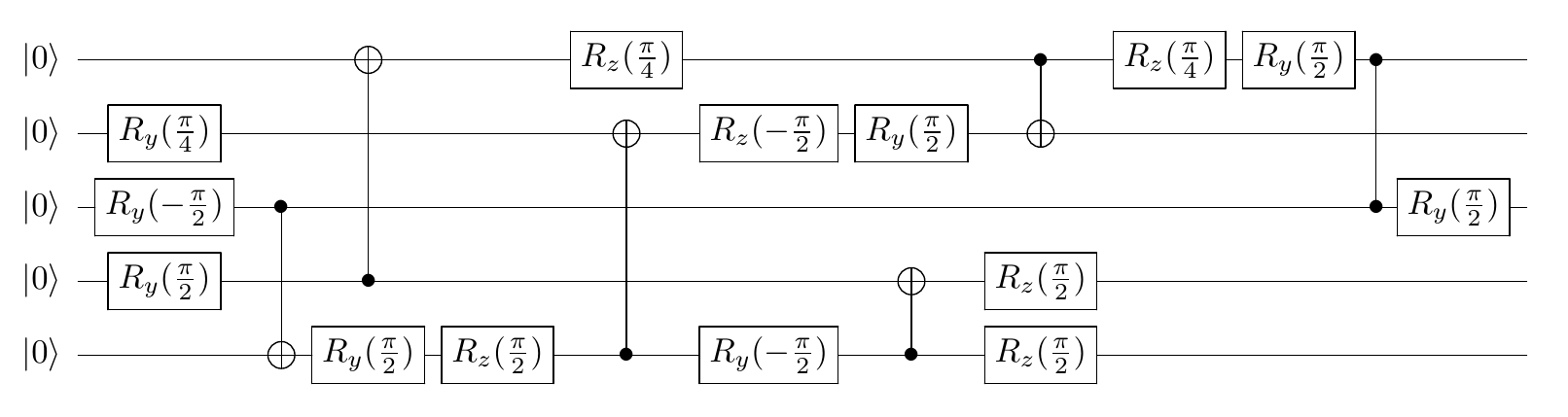}\\
(b)\\
\includegraphics[width=2\columnwidth]{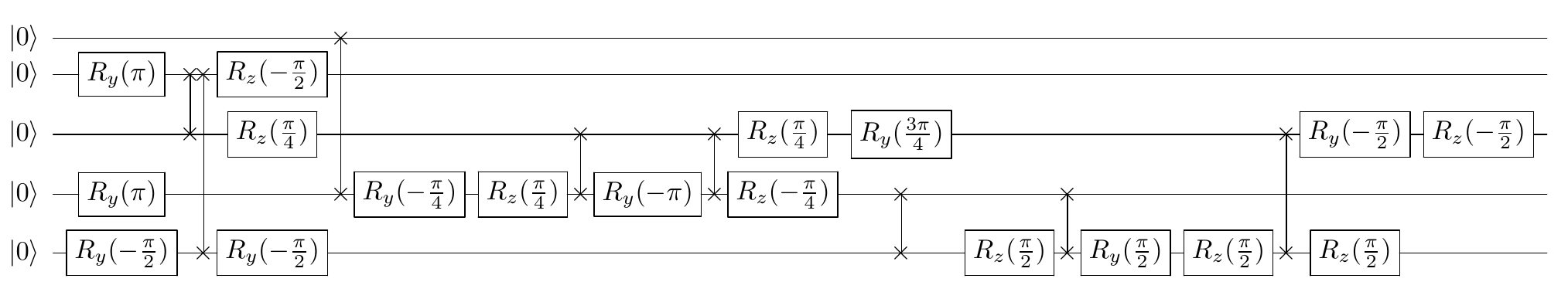}\\
(c)\\
\includegraphics[width=2\columnwidth]{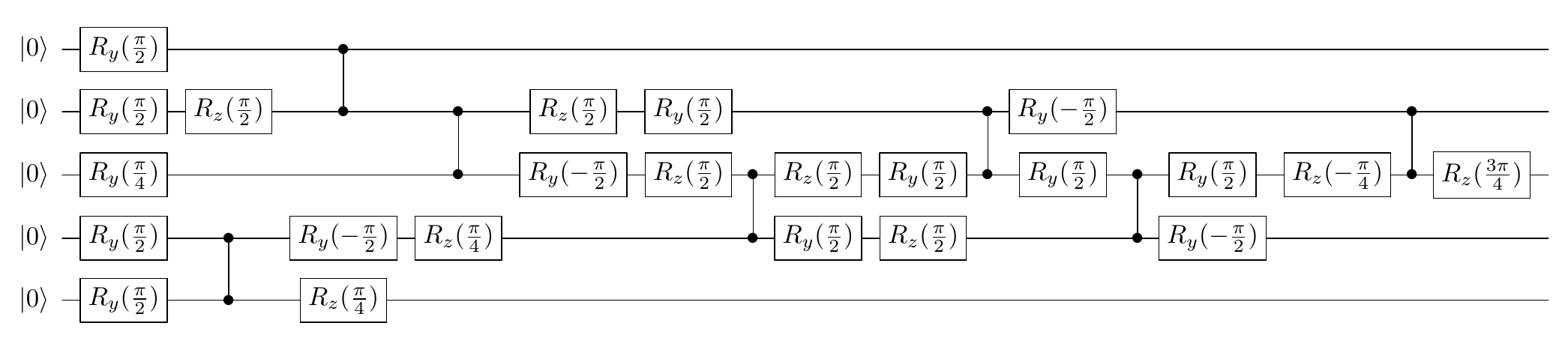}\\
(d)\\
\caption{
(a) Circuit to encode a logical minus state $\ket{-}_L$, with the five qubit code. This circuit is equivalent to Fig.~\ref{fig:five-minus}(a).
(b) Circuit to encode a logical magic state $\ket{T}_L=\frac{\sqrt{2}}{2}(\ket{0}_L+e^{\frac{\pi}{4}i}\ket{1}_L)$, with the five qubit code. Our method found at minimum six two-qubit gates are required to prepare the state.
(c) Circuit to encode a logical minus state $\ket{-}_L$ with the five qubit code. The two-qubit gate is restricted to be a sqrt-SWAP gate.
(d) Circuit to encode a logical magic state $\ket{T}_L=\frac{\sqrt{2}}{2}(\ket{0}_L+e^{\frac{\pi}{4}i}\ket{1}_L)$. Only nearest-neighbour CPhase gates are permitted in this case.
}
\label{fig:five-magic}
\end{centering}
\end{figure*}

\begin{figure*}
\begin{centering}
\includegraphics[width=1.4\columnwidth]{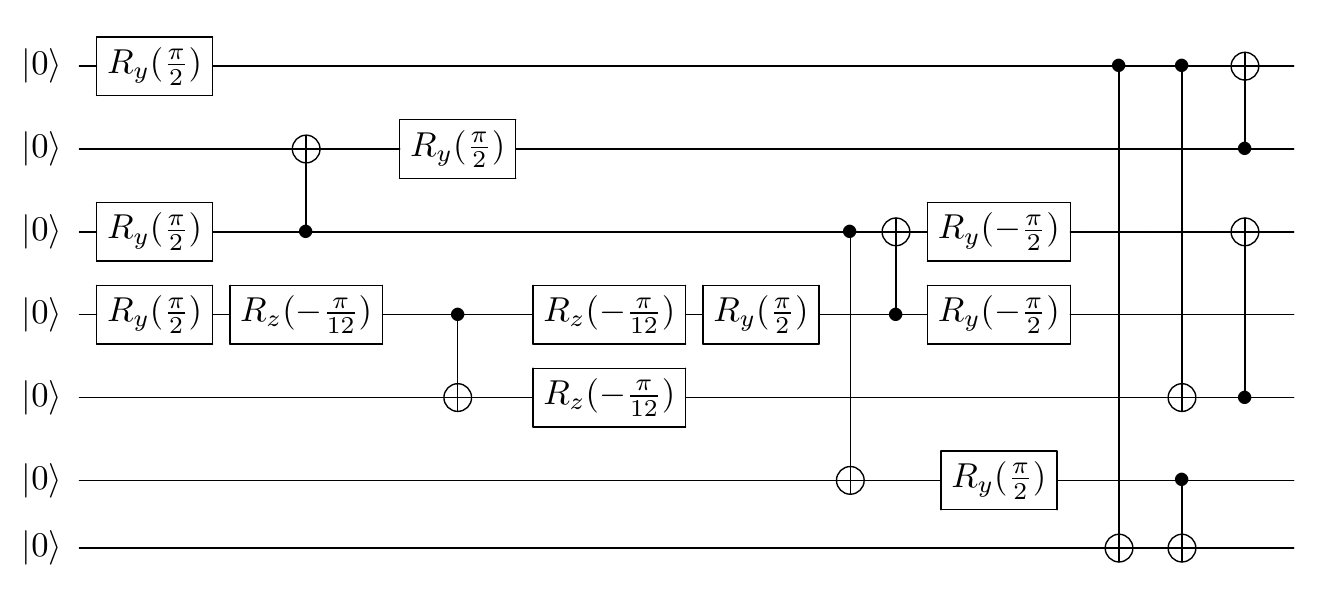}\\
\caption{Circuit to encode a logical magic state $\ket{T}_L=\frac{\sqrt{2}}{2}(\ket{0}_L+e^{\frac{\pi}{4}i}\ket{1}_L)$ with the seven qubit code. Minimally 9 two-qubit gates are required to prepare the state.}
\label{fig:seven-magic}
\end{centering}
\end{figure*}

\subsection{Compiling with ideal gates}
We first consider the case where gates are assumed to be perfect. In this case, we can focus on the optimisation with pure state imaginary time evolution. 

We begin with the five qubit code. By considering the circuit with the minimum number of two-qubit gates, we first rediscover circuits for encoding the $\ket{-}_L$ state, which are consistent with the latest conventional circuit as shown in Fig.~\ref{fig:five-minus}(a). We also note that with five controlled phase gates, the circuit is capable of encoding the $\ket{0}_L$ state with additional transversal single-qubit gates. As our method is capable of discovering different but equivalent circuits, we show one example in Fig.~\ref{fig:five-magic}(a), which encodes a logical minus state $\ket{-}_L$.

Next, we apply our compiler for the magic state $\ket{T}_L$ and we find the encoding circuit as shown in Fig.~\ref{fig:five-magic}(b). The encoding circuit for the magic state is also consistent with the circuit for encoding an arbitrary logical state as shown in Fig.~\ref{fig:five-minus}(b). {Therefore, our method has matched but not surpassed the efficiency of a known circuit for encoding the magic state.}

In addition to rediscovering existing encoding circuits, our compiler can also find efficient encoding methods when considering a variety of constraints on the circuit structure. First, we consider the case where sqrt-SWAP gates are considered as the only type of two-qubit gates in the ansatz. The sqrt-SWAP gate is a non-Clifford gate, so it is not often seen in conventional error correction encoding circuits. On the other hand, it may be a natural two-qubit gate~\cite{alonso_calafell_quantum_2019}. The canonical approach would be to  convert this gate into a CNOT/CPhase gate in a circuit design. As one CNOT gate is decomposed into two sqrt-SWAP gates and several single-qubit rotations, at minimum 10 sqrt-SWAP gates are then required to encode a $\ket{-}_L$ state. 
We show here that by applying the sqrt-SWAP gate into the ansatz directly, the number of the sqrt-SWAP gates can be reduced to eight as shown in Fig.~\ref{fig:five-magic}(c). 
Next, we consider the case where the ansatz is restricted to allow only nearest-neighbour interactions, which is common for solid state qubits. We present in Fig.~\ref{fig:five-magic}(d) a circuit satisfying the constraint, which prepares the magic state $\ket{T}_L$ with seven nearest-neighbour CPhase gates. (Note that here we specified that CPhase gates should be the only two-qubit gates; if we relax this and subsume certain single-qubit and CPhase gates into CNOT gates, then this circuit simplifies further.)

For the seven qubit Steane code, we also first rediscover the encoding circuits for the $\ket{0}_L$ state as shown in Fig.~\ref{fig:seven-zero}(a), which has 8 CNOT gates. For the magic state, we find a circuit that only uses 9 CNOT gates, in contrast to the encoding circuit for an arbitrary state, which requires 11 CNOT gates as shown in Fig.~\ref{fig:seven-zero}(c).

\subsection{Noise-robust circuits}
In this section, we consider the practical scenario with noisy gates. For each gate, we apply a small probability of depolarising noise as follows,
\begin{equation}
\begin{aligned}
\rho^\prime\to(1-r)\rho+\frac{r}{3}(X\rho X+Y\rho Y+Z\rho Z).
\end{aligned}
\end{equation}
We emphasise that {\em any} {theoretically given noise model in a simulator or any practical noise model in physical hardware} could be employed here; we consider depolarising noise for this first study in order to more readily compare to known methods and results. 
Then we aim to find the most `noise-robust' encoding circuit for preparing a target logical state. Here, we also consider an ancillary qubit, which is applied to the circuit and post-selected in a similar way to  fault-tolerant state preparation. Note that when the ancilla is not entangled with the physical qubits, it reduces to the previous case where there is no ancilla. As the encoded state is a mixed state, we make use of the mixed state mode of the QuEST and the imaginary time evolution for mixed states. 
We search over a large number (order 10000) of different ans\"atze, and we obtain the circuit corresponding to the lowest energy. We then verify that this circuit performs better in the presence of small gate noise than the circuits found in the earlier section, which were found under the zero-noise assumption.

\begin{figure}
\begin{centering}
\includegraphics[width=1\columnwidth]{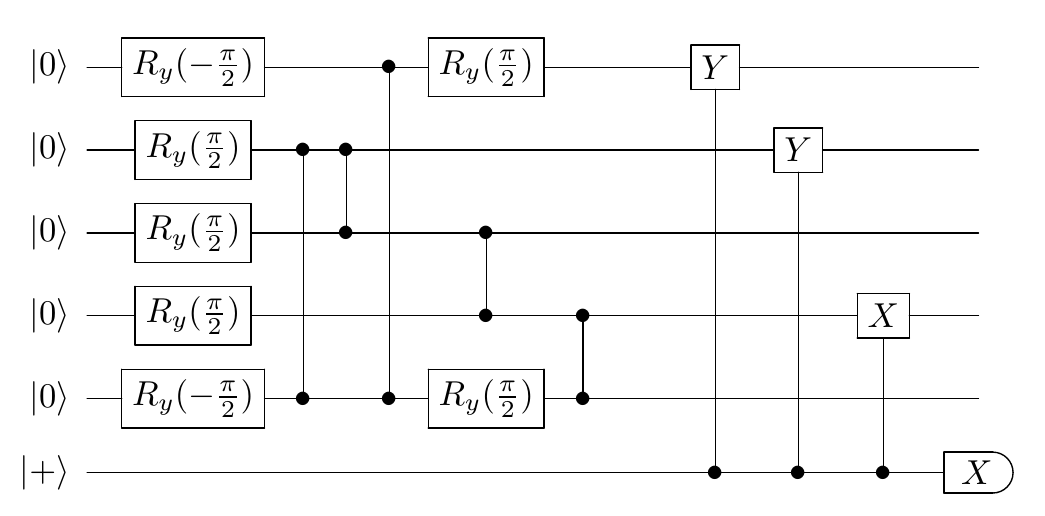}\\
\caption{A noise-robust circuit to encode a logical minus state $\ket{-}_L$ with the five qubit code. In addition to the first part of the circuit (Fig.~\ref{fig:five-magic}(a)) which prepares $\ket{-}_L$, three two-qubit gates are applied from the ancilla qubit, which is then measured in the $\{+,-\}$ basis. If measured to be $-$, the output is abandoned and re-prepared until the measurement result is $+$.}
\label{fig:fiveFT}
\end{centering}
\end{figure}

For the five qubit code, a noise-robust circuit is found for encoding the logical minus state $\ket{-}_L$ as shown in Fig.~\ref{fig:fiveFT}. 
The circuit contains two parts, with the first part (gates on the data qubits) preparing a logical minus state, while the second part (gates between the data qubits and the ancilla) detecting and post-selecting errors. Note that the second part is a logical $X_L$ operator which does not change the logical state. 

\subsection{{Error detection for general encoding circuits}}

It is interesting to reflect further on the principle of the circuit which our automated method has found. We could generalise the rule to any error correction codes with transversal Pauli gates. For example, a noise-robust circuit to encode any of the logical states $\ket{0}_L,\ket{1}_L,\ket{+}_L,\ket{-}_L,\ket{+i}_L,\ket{-i}_L$ could be realised by preparing the states with the non-fault-tolerant (non-FT) circuits first and applying a transversal logical operator (which does not change the logical state) for detecting and post-selecting errors. We notice that by combining the transversal logical operator with the stabiliser operator, some Pauli terms can be cancelled out. If replaced with this new operator for error detection, the circuit is noise-robust. Note that the circuit shown in Fig.~\ref{fig:fiveFT} is not fully fault tolerant, as the logical state prepared with the five qubit code can be corrupted by any single error, while measuring out one logical operator and applying post selection cannot guarantee FT detection of any single error. However, circuits to prepare some logical states with the Steane code applied with error detection can be realised fault-tolerantly, as the error detection needs only to detect a certain type of noise, while the logical states are immune to the other type of noise. For example, a logical $\ket{0}_L$ encoded with the Steane code is immune to any phase noise. The circuit discovered by Goto~\cite{goto2016minimizing} to prepare a FT logical zero state $\ket{0}_L$ with the Steane code (as shown in Fig.~\ref{fig:seven-zero}(b)) is an example.

For the Steane code, one can implement transversal Clifford gates, so one may feel that the same trick can be applied. That is, to test the fault tolerance of the circuit, we measure the logical operator $P$, when it belongs to the Clifford group and satisfies $P\ket{\Phi}_L=\ket{\Phi}_L$, with $\ket{\Phi}_L$ being the target logical state. Unfortunately, if an error occurs to a data qubit, an extra gate needs to be applied to the data qubit even if the ancilla is measured to be $+$. Therefore, a subsequent error detection or error correction procedure is still required to remove the single error, as pointed out in Ref.~\cite{goto2016minimizing,chamberland2019fault}.

\begin{figure}
\begin{centering}
\includegraphics[width=0.6\columnwidth]{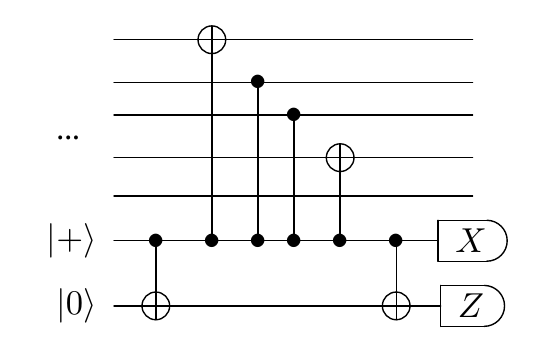}\\
\caption{Circuit to fault-tolerantly measure one stabiliser operator with the five qubit code. The first and second ancillae need to be measured to be + and 0, or otherwise the output is abandoned.}
\label{fig:fiveStabiliser}
\end{centering}
\end{figure}

\begin{figure}
\begin{centering}
\includegraphics[width=0.98\columnwidth]{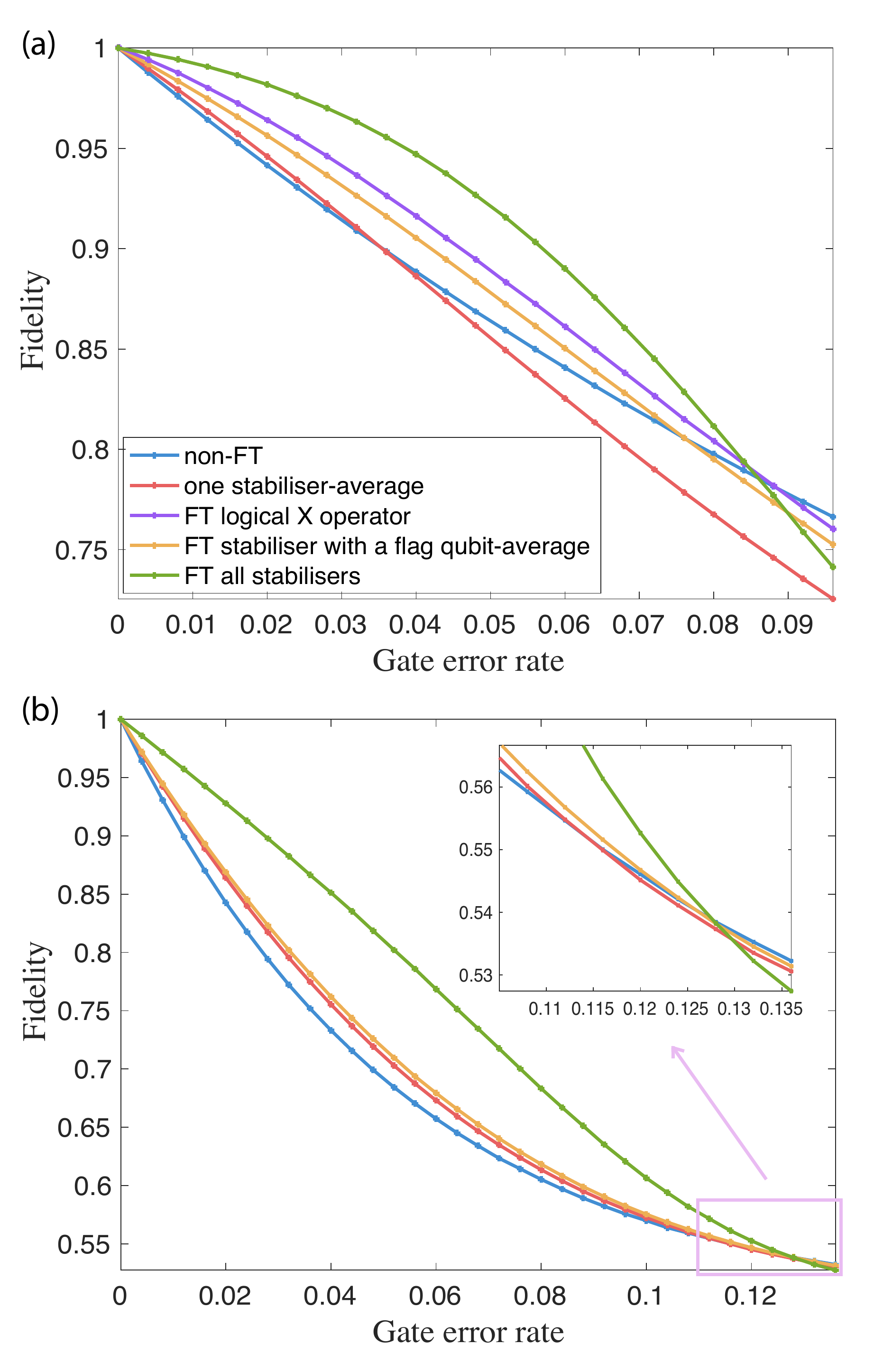}\\
\caption{The fidelity change with the increasing gate error rate in preparation of (a) a logical minus state $\ket{-}_L$ with the five qubit code and (b) a logical magic state $\ket{T}_L=\frac{\sqrt{2}}{2}(\ket{0}_L+e^{\frac{\pi}{4}i}\ket{1}_L)$ with the seven qubit code. The error rate for single-qubit gates, two-qubit gates and measurements is the same. The blue curves represent the case where the logical state is prepared with the non-FT circuits in Fig.~\ref{fig:five-magic}(a) and Fig.~\ref{fig:seven-magic} for the two codes respectively. If one of the four/six stabiliser operators is measured with one ancilla and post selection is applied, the average behaviour of the four/six cases is shown as the red curves, while the orange ones refer to the same scenario but the operators are measured fault-tolerantly with two ancillae. The purple curve in (a) refers to the case where a logical X operator is measured, with the circuit shown in Fig.~\ref{fig:fiveFT}, except that the error detection is conducted fault tolerantly with additional ancilla. The greens refer to the case where all the stabiliser operators are measured with two ancillae.}
\label{fig:magicSnttate-noiseReb}
\end{centering}
\end{figure}

In general, there are no universal transversal logical operators for small error correcting codes. Thus we cannot apply the same trick to fault tolerantly prepare an arbitrary logical state, such as the magic state of the five and seven qubit code. However, a FT circuit can be realised by measuring all the stabiliser operators and applying post selections: the ancilla should always be measured to be 1, otherwise the circuit is abandoned and restarted from the beginning. Note that measuring all the stabiliser operators cannot guarantee fault tolerance of the circuit -- a logical error created before the error detection cannot be found. As it takes a relatively long time to prepare a logical state, this approach may not be ideal for systems with short coherence time. On the other hand, one can measure one of the stabiliser operators and still benefit partially.

We also note that measuring the stabiliser operators can also be fault tolerant with an extra flag qubit, as shown in Fig.~\ref{fig:fiveStabiliser}. After a logical state is prepared non-fault-tolerantly with the five qubit code, we measure one of the stabiliser operators. The flag qubit is to detect errors occurring between one of the two-qubit gates as shown in the figure.  

Finally, we compare the fidelity of the prepared logical state with different encoding circuits as discussed above.  
In Fig.~\ref{fig:magicSnttate-noiseReb}, we show the fidelity change with respect to an increasing gate error rate. The error rate is the same for single-qubit gates, two-qubit gates and measurement {and noise is applied for both the state preparation and error detection processes.} A logical minus state is prepared with the five qubit code with circuit shown in Fig.~\ref{fig:five-magic}(a). We see that with small noise, applying error detection always leads to a higher fidelity. However, as the gate error rate gradually increases, the extra noise introduced with the extra gates negates the advantage of error detection. In Fig.~\ref{fig:magicSnttate-noiseReb}(a), there is a small gap between the blue and the red curves, indicating that measuring one stabiliser operator non-fault-tolerantly only gains a small benefit over the case where no error detection is performed. However, such an advantage increases if the measurement is conducted fault-tolerantly with one more ancilla. The purple curve, representing post selection by measuring the X operator, has an overall higher fidelity than the red and orange ones probably due to one fewer two-qubit gate applied. If all the stabiliser operators are measured fault-tolerantly with post selection applied, as demonstrated by the green curve, the fidelity is higher than the non-FT circuit when the gate error rate is smaller than 8.6\%. Note we have verified that in this case, the circuit involving measuring all the stabilisers fault-tolerantly is fault-tolerant. In Fig.~\ref{fig:magicSnttate-noiseReb}(b), a logical magic state $\ket{T}_L=\frac{1}{\sqrt{2}}(\ket{0}_L+e^{\frac{\pi}{4}i}\ket{1}_L)$ is prepared with the seven qubit code. Compared with (a), we see a group of curves with different shapes but a similar trend, that the curves applied with post selection have a higher fidelity given a small gate error rate. The advantage starts to vanish when the gate error rate is larger than 11\%. In this case, we found the circuit involving measuring the full stabiliser set is not fault-tolerant but still leads to a notably higher state fidelity. The result suggests the noise-robustness of our method.

\section{Discussion}
\label{sec:discussion}
In this work, we introduce a variational circuit compiler for efficiently encoding the logical state of an error correcting code. We construct a Hamiltonian so that the target logical state is its ground state and it can be found with the variational imaginary time evolution method. We consider the five and seven qubit codes as examples. When having noiseless operations, we show the encoding circuit to prepare different logical states with the minimal number of gates for different hardware structures. When considering noisy gates, it is discovered by applying post selection, the fidelity of the encoded state can be boosted. Therefore we can simplify the compiling process by searching for more optimal non-FT circuit first with given constraints and noise models, and applying post selection after the state preparation process.

We also compare the fidelity of logical states prepared with different encoding circuits with respect to different gate error rates. Our work thus opens an avenue for automatically compiling circuits for implementing error correcting codes. Future studies may focus on the design and searching of ans\"atze for different codes and the realisation of the compiler with a real quantum computer. 

It is natural that some of the highest-performing circuits found by our approach were previously known, since we opted to explore two very well-studied codes and moreover we employed a noise model (depolarising noise) which has been the canonical model of choice for previous work. {However we emphasise that our approach is by no means limited to these choices and can be used to optimise circuits for newly emerging codes -- it is even practical to seek optimisation of larger codes, as we can start from the previously-discovered circuit and search for better ones, based on bespoke error models that are matched to specific hardware implementations.} 

\section{Acknowledgements}
SCB and XX are supported by the Office of the Director of National Intelligence (ODNI), Intelligence Advanced Research Projects Activity (IARPA), via the U.S. Army Research Office Grant No. W911NF-16-1-0070. The views and conclusions contained herein are those of the authors and should not be interpreted as necessarily representing the official policies or endorsements, either expressed or implied, of the ODNI, IARPA, or the U.S. Government. The U.S. Government is authorized to reproduce and distribute reprints for Governmental purposes notwithstanding any copyright annotation thereon. Any opinions, findings, and conclusions or recommendations expressed in this material are those of the author(s) and do not necessarily reflect the view of the U.S. Army Research Office. SCB and XY acknowledge EPSRC grant EP/M013243/1. SCB acknowledges support from the European Quantum Technology Flagship project AQTION. The authors would like to acknowledge the use of the University of Oxford Advanced Research Computing (ARC) facility in carrying out this work. http://dx.doi.org/10.5281/zenodo.22558

\bibliography{circuit.bib}

\appendix

\section{Proof of Eq.~\eqref{Eq:fidelity}}
Here we prove Eq.~\eqref{Eq:fidelity}. Suppose we have a Hamiltonian $H$ with ground and first excited energy denoted by $E_0$ and $E_1$, respectively. Given a quantum state $\rho$ with averaged energy $E=\tr[\rho H]$ satisfying $E\in [E_0, E_1]$, we want to prove 
\begin{equation}\label{Eq:fidelityApp}
\begin{aligned}
F = {\braket{\psi_0|\rho|\psi_0}} \ge 1-(E-E_0)/c,
\end{aligned}
\end{equation}
where we denote $c=E_1-E_0$ and $\ket{\psi_0}$ being the ground state of $H$. 

\begin{proof}
Denote the eigenstates of $H$ by $\ket{\psi_i}$ with corresponding eigenvalues $E_i$ satisfying $E_i\le E_j$ when $i\le j$. Then we have 
$$H=\sum_i E_i \ket{\psi_i}\bra{\psi_i},$$
and 
\begin{equation}
    \begin{aligned}
    E=&\tr[\rho H],\\
    =&\sum_i E_i \bra{\psi_i}\rho\ket{\psi_i},\\
    =&E_0F+\sum_{i\ge1} E_i \bra{\psi_i}\rho\ket{\psi_i},\\
    \ge&E_0F+\sum_{i\ge1} E_1 \bra{\psi_i}\rho\ket{\psi_i},\\
    =& E_0F + E_1 (1-F),\\
    =& (E_0+c)-cF.
    \end{aligned}
\end{equation}
In the third line, we make use of the definition of $F$ in Eq.~\eqref{Eq:fidelityApp}; In the fourth line, we make use of the ordering of the eigenvalues; In the fifth line, we make use of $\sum_{i\ge1} \bra{\psi_i}\rho\ket{\psi_i}=1-\bra{\psi_0}\rho\ket{\psi_0}=1-F$.
Solving the equation, we thus get the lower bound of the fidelity $F$ based on the energy $E$ as in Eq.~\eqref{Eq:fidelityApp}.
\end{proof}

\end{document}